\documentclass[aps,prl,superscriptaddress,amssymb,twocolumn,showpacs,a4]{revtex4}

\usepackage{epsfig}
\usepackage{graphics}
\usepackage{graphicx}
\usepackage{amsmath}

\begin{document}

\title{Scaling of dynamics in 2d semi-dilute polymer solutions}

\author{Pietro Cicuta}
\affiliation{Cavendish Laboratory, University of Cambridge,
Madingley Road, Cambridge CB3 0HE, U.K.}
 \email[]{pc245@cam.ac.uk}
\author{Ian Hopkinson}
\affiliation{Department of Physics, U.M.I.S.T., Manchester M60
1QD, U.K.}

\begin{abstract}
 We study the viscoelasticity of surface polymer monolayers by
 measuring the dynamics of thermal concentration fluctuations with surface light scattering.
For various systems of proteins and synthetic polymers we find a
 semi-dilute regime in which both the elastic and viscous
components of the dilational modulus  show power law dependencies
on the concentration. Surprisingly there is a universal
relationship between the  exponents for the two components: the
viscosity scales with a power double that of the elasticity. These
results cannot be explained on the basis of theory developed for
bulk systems, and a simple explanation for the singular 2d
behavior is suggested.
\end{abstract}

\pacs{{68.03.Cd}, {87.14.Ee}}

\maketitle Polymers are flexible long chain molecules, of
outstanding importance in diverse fields from processing of
materials to biological activity.  It is well known that it is
possible to effectively confine some polymers to two dimensions,
for example by anchoring each monomer to the interface between
immiscible fluids~\cite{gaines66}. Polymer dynamics in two
dimensions (2d) has remained relatively unexplored compared to
bulk solutions, despite there being important examples both in
life sciences and in technology where polymer molecules are
confined to a plane. We consider the simplest situation where the
chains lie flat and are free to move only in a 2d space.
 Such a surface layer exerts a
lateral osmotic pressure $\Pi=\gamma_0-\gamma$, which is the
amount by which the surface tension $\gamma$ is reduced compared
to the free interface tension $\gamma_0$. This pressure becomes
significant above overlap of single chains, where it has a
power-law dependence  on the concentration that has been known for
some time~\cite{rondelez80}. Like its counterpart in 3d bulk
solutions, this equilibrium property is well explained within the
picture of the polymer semi-dilute regime as an ensemble of
independent ``blobs''~\cite{deGennes79}. Despite a few decades of
experiments, very little is known instead about the physical
nature of the divergence of viscosity and elasticity in 2d. These
dynamical  properties control important processes like foam
drainage~\cite{stone03} and stabilization and flow in
emulsions~\cite{buzza95} when polymers are used as surfactants.
The theoretical framework to describe complex non-Newtonian flow
and the dynamics of chains is well understood in
3d~\cite{edwards86}, but cannot be applied straightforwardly in
2d. How does the confinement to the surface, specifically a 2d
effect, affect the dynamics of the system?

 This Letter presents a study of the dynamics of concentration fluctuations
 in a wide range of polymer monolayers, performed with surface dynamic light scattering.
Polymers  are spread onto an interface in dilute conditions, and
cannot subsequently submerge and re-surface. Under these
conditions polymer chains are unable to cross each other.
This is the case for many monolayer systems studied in the
literature. The technique used in this work, developed by Langevin
and others, see the monograph~\cite{lan92}, is the only existing
probe of thermal concentration fluctuations in monolayers.
Despite its use on polymer monolayers by various
groups~\cite{various},
   some issues concerning the data analysis have been resolved only recently,
hence the data which is available for a quantitative comparison is
very limited.

\begin{figure}[t]
         \epsfig{file=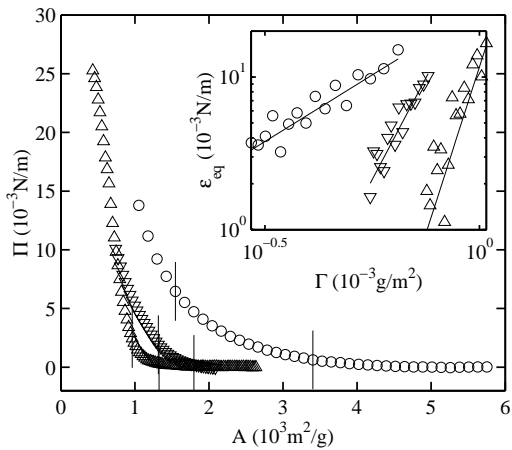,height=6cm}
    \caption{Surface pressure  $\Pi$ as a function of area. Data
    are
representative of three systems studied in this Letter:
($\circ$)~PVAc, ($\triangledown$)~$\beta$-casein,
($\vartriangle$)~$\beta$-lactoglobulin. Segments indicate the
semi-dilute regime region where the equilibrium scaling exponents
are determined. Inset:~log plot of the equilibrium dilational
modulus $\varepsilon_{eq}$ against concentration. Lines are fitted
power laws, showing  the well known scaling of this equilibrium
property in the semi-dilute regime. \label{figure1}}
\end{figure}

 Isolated polymer chains are usually modelled
as random walks with potential interactions~\cite{deGennes79}. The
Flory exponent $\nu$ relates the number of monomers $N$ to the
radius of gyration: $R_g \sim N^\nu$. When the concentration in a
polymer solution is increased so that individual chains are forced
to overlap, the system enters a semi-dilute regime that lasts
until the monomer fraction is very high. The semi-dilute region is
important because for sufficiently long chains it covers a wide
range of concentrations. The equilibrium properties of polymers in
this regime are  given  both in 2d and 3d  by scaling laws, with
exponents related to $\nu$. In particular, fluctuations of the
density of monomers are correlated over a length $\xi$, defining a
region known as a ``blob"~\cite{deGennes79}. Inside the blob the
chain has the same statistics as an isolated single chain, but
different blobs are statistically independent. In 2d $\xi$ scales
with the concentration $\Gamma$ as:
  \begin{equation}
\xi\sim\Gamma^{\nu/(1-2\nu)}, \label{eq1}
  \end{equation}
decreasing from a value of the order of $R_g$ at the overlap
concentration to the monomer size. Scaling of the osmotic pressure
in a monolayer was first shown experimentally in
ref.~\cite{rondelez80}:
\begin{equation}
\Pi \sim \xi^{-2}\sim \Gamma^{y_{eq}},\,\, {\textrm{where}}\,\,
y_{eq}=2\nu/(2\nu-1). \label{eq2}
  \end{equation}
    Figure~\ref{figure1} shows
equilibrium data for some of the systems studied in this work. In
2d there are well known limiting regimes: The ``good solvent", in
which the chain behaves as a self-avoiding random walk, hence
$\nu=3/4$ and $y_{eq}=3$, and the ``$\theta$~conditions" where the
effects of excluded volume balance the monomer-monomer
preferential attraction. Here the Flory exponent is predicted to
be $\nu=4/7$~\cite{coniglio87} and the power law exponent
increases to $y_{eq}=8$, making this isotherm much steeper and
compact. In contrast to the bulk, where the polymer chains can
interpenetrate, in the planar 2d geometry the polymers are
segregated and attain disk-like configurations~\cite{deGennes79}.
Chains are corralled by their neighbors, as was observed with DNA
molecules confined to a plane in ref.~\cite{radler99}.

\begin{figure}[t]
         \epsfig{file=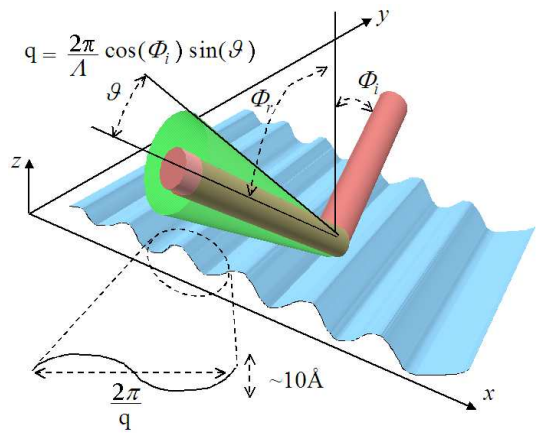,height=6cm}
    \caption{(color online) The geometry of scattering from thermal surface roughness
fluctuations, as probed by surface light scattering (SQELS).
Typical values of the scattering vector $q$ are of the order of
400cm$^{-1}$, corresponding to capillary wave frequencies $\omega$
of order 10$^5$Hz. $\Lambda$ is the frequency of the laser light.
Details of the technique are given in
ref.~\cite{cicuta03b}.\label{figure2}}
\end{figure}

 In general,
the response to a deformation in an isotropic 2d material is
characterized by two elastic moduli: changes in area are
controlled by the dilation modulus $\varepsilon$ and changes in
shape by the shear modulus $G$~\cite{miller96}. Polymer monolayers
in the semi-dilute regime are fluid-like and the shear modulus is
negligible~\cite{cicuta03}. In these conditions it is common
practice to determine the equilibrium dilational modulus
$\varepsilon_{eq}$ (the 2d analog of the bulk modulus) from
measurements of pressure as a function of area: $\varepsilon_{eq}
= d\Pi/d \ln A$. By definition, $\varepsilon_{eq}$ has the same
scaling properties as the osmotic pressure, described in
Eq.~\ref{eq2}. Power laws of $\varepsilon_{eq}$ as a function of
concentration are shown in Fig.~\ref{figure1}.
The dilational modulus can also be measured dynamically, in an
experiment where the surface area oscillates in time with a
frequency $\omega$. Then the dynamic complex modulus
$\varepsilon^*=\varepsilon'+ i\omega \varepsilon''$ is probed,
where $\varepsilon'$ is the elastic component of the response
modulus and  $\varepsilon''$ is the dilational viscosity.
$\varepsilon^*$ can be accessed with the surface quasi-elastic
light scattering (SQELS) technique, shown in Fig.~\ref{figure2}.
SQELS measures the time correlation function of light scattered
from thermal surface roughness, which acts as a phase
grating~\cite{lan92}. These out of plane fluctuations are
underdamped waves with a frequency $\omega$, and their motion is
affected by the presence of a surface film.  To recover surface
viscoelaticity, the surface wave dispersion relation $D(\omega)$
that relates the wave frequency to the wavelength has to be known:
 \begin{align}
D(\omega)&=\left[\varepsilon^* q^2+i\omega \eta \left( q+m\right)
\right]\left[\gamma q^2+i\omega\eta \left( q+m \right)-\frac{\rho
\omega^2}{q} \right]+\nonumber
\\&-\left[i\omega \eta \left(m-q \right) \right]^2,
\textrm{ where } m\,=\,\sqrt{\,q^2\,+\,i\frac{\omega
\rho}{\eta}}\nonumber\\ &\textrm{ and  } {\rm Re}(m)>0,
 \label{eq5}
\end{align}
$\eta$ is the subphase Newtonian viscosity and $\rho$ is the
subphase density. Buzza~\cite{buzza02} recently
 proved that the model Eq.~\ref{eq5}, which is commonly found in the literature~\cite{lucassen69,lan92}, is correct but
 that even
under dynamical conditions the surface tension $\gamma$ should be
considered as a real quantity, equal to the equilibrium static
surface tension~\footnote{
 The model often used in the literature is over-parametrized. Data analysis
 with the correct physical parameters is very robust.}. As derived in
ref.~\cite{langevin71}, the spectrum of light scattered by thermal
roughness is given by:
\begin{eqnarray}
P_q(\omega)\,=\,\frac{k_B T}{\pi \omega}{\rm Im}\left[\,\frac{i
\omega \eta (m\,+\,q)\,+\,\varepsilon q^2 }{D(\omega)}\, \right].
 \label{eq6}
\end{eqnarray}
SQELS data is fitted with Eq.~\ref{eq5} and Eq.~\ref{eq6}, with
only three physical parameters to be determined. Details of our
experimental methods, including calibration, data analysis and
limits of the technique, are described in ref.~\cite{cicuta03b}.

 In polymer systems, different
models are appropriate depending on the lengthscales and
timescales that are being observed. The lengthscales $q^{-1}$
probed with SQELS satisfy $q\xi<1$, so the semidilute solution is
expected to be in a ``macroscopic'' regime where it behaves like a
gel~\cite{degennes76}. The gel's response is characterized by a
rigidity modulus, $\varepsilon'$, scaling like the number of
contacts, hence proportional to the osmotic pressure, and by a
viscous dissipation modulus $\omega\varepsilon''$ describing the
friction involving the monomers and the solvent. This is well
established  in three dimensional solutions, where both a fast and
a slow relaxation mode can be probed simultaneously, for example
with dynamic light scattering~\cite{brown93}. The fast relaxation
can be described within the Rouse model and is related to a
cooperative diffusion timescale $\tau_{coop}$~\cite{degennes76}.
This describes fluctuations that become faster as the
concentration increases and the correlation length $\xi$
decreases. The slow relaxation is related to the self diffusion
coefficient for reptation and is determined by the time
$\tau_{rept}$ required for the chain to diffuse along its length.
$\tau_{rept}$ is an increasing function of the concentration. The
approach of  directly measuring  the frequency spectrum of light
scattered by concentration fluctuations  is not possible in
monolayers because there is insufficient optical contrast, and
dynamics can be probed only by indirect methods such as SQELS.

\begin{figure}
         \epsfig{file=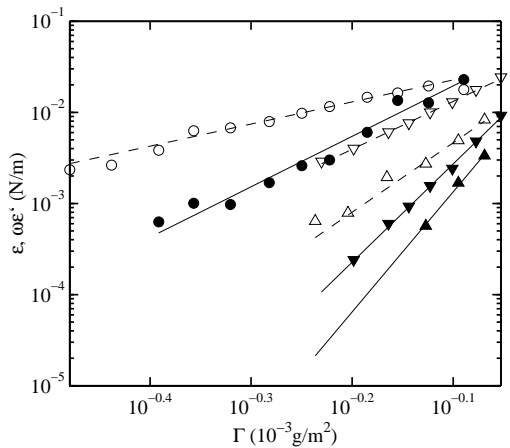,height=6cm}
    \caption{Log plot of
the dilational elastic response modulus measured with SQELS as a
function of the concentration, showing power-law scaling. Open
symbols are the elastic ($\varepsilon'$)  and filled symbols the
viscous ($\omega \varepsilon''$) components. Symbols are for
different monolayers: ($\circ$)~PVAc,
($\triangledown$)~$\beta$-casein,
($\vartriangle$)~$\beta$-lactoglobulin, corresponding to data in
Fig.~\ref{figure1}. Standard deviations obtained from fitting
repeated SQELS correlation functions (not shown on each data point
for clarity) are used to weigh the power-law interpolation
(lines). The exponents obtained from these fits are given in
Table~\ref{table} and are plotted in Fig.~\ref{figure4}.  The
correlation between the different exponents, highlighted in
Fig.~\ref{figure4}, represents the major result of this Letter.
\label{figure3}}
\end{figure}

Figure~\ref{figure3} shows the components of the monolayer
viscoelatic moduli, measured with SQELS for  different polymer
monolayers. Both the elastic and viscous components of
$\varepsilon^*$ exhibit a power law dependence on the
concentration:
\begin{equation}
\varepsilon'\sim \Gamma ^{y_{\varepsilon'}} \,\,\textrm{and}\,\,
\varepsilon''\sim \Gamma ^{y_{\varepsilon''}}  \label{eq3}
  \end{equation}
    A similar scaling
behavior was recently reported by Monroy {\it et
al.}~\cite{monroy99} but has not been explained~\footnote{An
attempt in ref.~\cite{monroy99} to understand the scaling of
$\varepsilon''$ within the Rouse model predictions for
$\tau_{coop}$ gave a dimensionally wrong result, which clearly has
no physical meaning.}. Results for all the systems considered in
this work are summarized in Table~\ref{table}. Monolayers of very
different compositions, comprising both synthetic polymers and
proteins, have been studied, to  cover the widest possible range
of values of $\nu$. The temperature, ionic strength and $p$H of
the liquid subphase are all controlled, as they affect the chain
configuration at the surface.

\begin{figure}[t]
  \epsfig{file=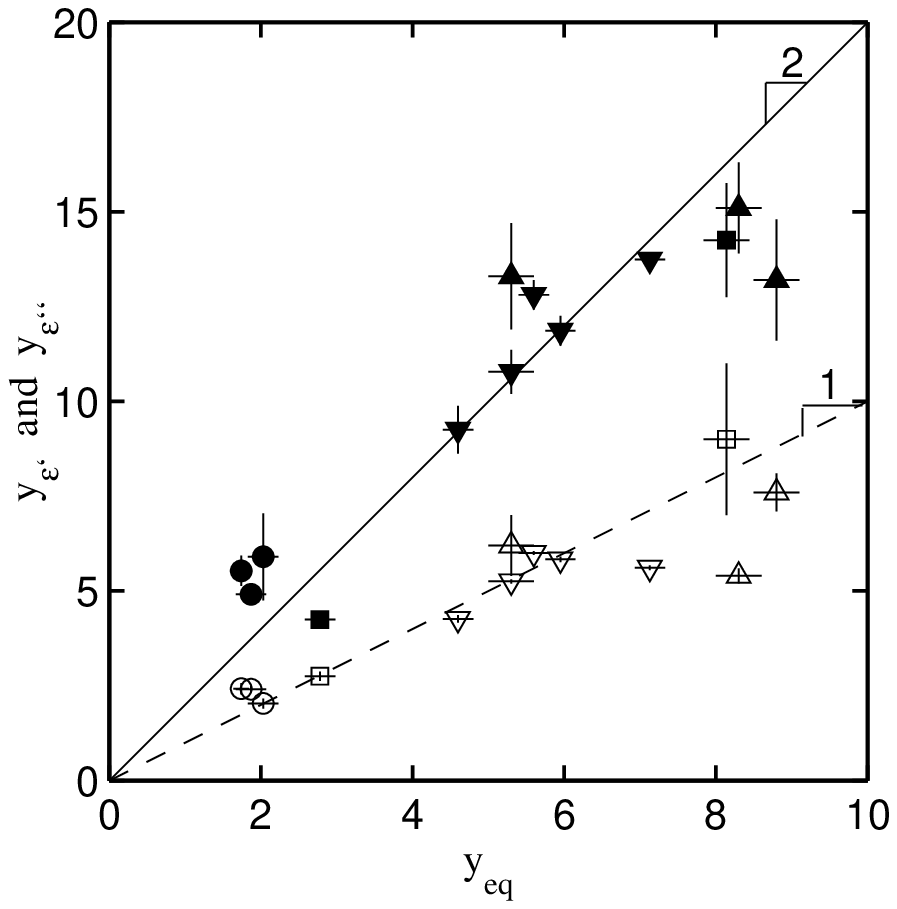,height=6cm}
    \caption{The scaling exponents describing the power law
dependence on the concentration of the dilational elasticity (open
symbols) and viscosity (filled symbols) are plotted against the
exponent for the equilibrium bulk modulus.  ($\bullet$):~PVAc,
($\blacktriangledown$):~$\beta$-casein,
($\blacktriangle$):~$\beta$-lactoglobulin, ($\blacksquare$):~data
of Monroy {\it et al.}~\cite{monroy99}. This figure includes
results from the data shown in Fig.~\ref{figure3} as well as the
monolayers described in Table~\ref{table}.  The lines have slopes
of 1 and 2. It is clear from the data that the exponent for the
viscosity is approximately twice the exponent of the elastic
modulus. This result is different from analogous measurements on
bulk polymer solutions, indicating a specific 2d dynamics.
\label{figure4}}
\end{figure}

This data contains a wealth of information.  Looking at
Fig.~\ref{figure3}, it can be clearly seen
 that for each monolayer the viscosity scales with a higher power
than the elastic modulus. On  general dimensionality terms, the
viscosity can be regarded as the product of the modulus driving
the relaxation and a characteristic
time~\cite{deGennes79}:~$\varepsilon''\sim \varepsilon' \cdot
\tau$. From this it immediately follows that the timescale $\tau$
of these relaxations becomes longer the higher the concentration,
hence a slow mode is being probed. The character of this mode can
be read from Figure~\ref{figure4},  where the power law exponents
for $\varepsilon^*$ defined in~Eq.~\ref{eq3} are plotted against
the equilibrium exponent. It is not surprising to find that
$y_{\varepsilon'}=y_{eq}$, since the thermal fluctuations probed
with SQELS fall within linear response theory. What is of interest
is the apparently universally valid relation between the exponents
for the elastic and the viscous components of $\varepsilon^*$:
$y_{\varepsilon''} = 2y_{\varepsilon'}$. It finally follows that
the timescale of these fluctuations, which we shall call
$\tau_{2d}$,
 scales like $\tau_{2d} \sim \xi^{-2}$.

\begin{table}
\begin{footnotesize}
\begin{tabular}{|c|c|c|c|c|c|}
  \hline
  Polymer & Conditions & Ref. & $y_{eq}$ & $y_{\varepsilon'}$ & $y_{\varepsilon''}$ \\
  \hline
PVAc &   T=45$^o$C & \cite{cicuta03b} &   2.0& 2.0& 5.9\\
PVAc &   T=25$^o$C & \cite{cicuta03b}  &  1.9& 2.4& 5.5\\
PVAc &   T=6$^o$C  & \cite{cicuta03b}  &  1.7& 2.4& 5.5\\
PVAc &   T=25$^o$C  & \cite{monroy99} & 2.8& 2.8& 4.3\\
$\beta$-casein &   $p$H=5.3, 0.01M & \cite{cicuta01} &7.1 &5.6 &13.7\\
$\beta$-casein & $p$H=7.2, 0.01M & \cite{cicuta01}   & 6.0& 5.8 &11.9\\
$\beta$-casein & $p$H=8.3, 0.001M & \cite{cicuta01} &5.6 &6.0 &12.8\\
$\beta$-casein &   $p$H=8.3, 0.01M & \cite{cicuta01} &   5.3 &5.3 &10.8\\
$\beta$-casein  &  $p$H=7.6, 1.1M   & \cite{cicuta01} &   4.6 &4.3& 9.3\\
$\beta$-lg & $p$H=6.0, 0.02M &    \cite{cicutaphd}         &   8.3 &5.4 &15.1\\
$\beta$-lg &   $p$H=8.3, 0.02M &  \cite{cicutaphd}         &     5.3& 6.2 &13.3\\
$\beta$-lg & $p$H=5.9, 0.1M&      \cite{cicutaphd}         & 8.8 &7.6 &13.2\\
P4HS & $p$H=2, T=25$^o$C &   \cite{monroy99}    &8.1 &9.0 &14.3\\
  \hline
\end{tabular}
\end{footnotesize}
\caption{Summary of static and dynamic scaling exponents for
monolayers studied in the literature. Values of $y_{eq}$ are
obtained from measurements of the equilibrium surface pressure
with Langmuir trough methods, while values of $y_{\varepsilon'}$
and $y_{\varepsilon''}$ are obtained from SQELS data. The SQELS
data of Cicuta and Hopkinson has been fitted following the method
described in ref.~\cite{cicuta03b}.
 The references specified  give experimental details for each monolayer. \label{table}}
\end{table}

 We are only able to
speculate on the physical origin of this timescale. A simple
explanation is that the slowing down of $\tau_{2d}$ as the
concentration increases is due to friction between the increased
number of statistically independent blobs. The number of contacts
in between blobs is proportional to $\xi^{-2}$, giving:
\begin{equation}\varepsilon''\sim\xi^{-4}\sim\Gamma^{2 y_{\varepsilon'}}
\label{eq4}
\end{equation}
which  describes  the data of Fig.~\ref{figure4} very well. An
analogy can be drawn between the diverging viscosity of the
close-packed arrangement of blobs in  the semi-dilute regime and
 the case of  diverging viscosity of
 spheres at high  packing density~\cite{brady93}, both being determined by the number of contacts.

A question that  arises is why this mode is not seen in 3d. In
bulk solutions in a  good solvent a dynamical mode analogous to
that just described for 2d would have a characteristic relaxation
time scaling with the concentration $\phi$ like $\tau\sim
\xi^{-3}\sim \phi^{9/4}$. This is a higher power of the
concentration compared to reptation ($\tau_{rept}\sim
\phi^{1.5}$)\cite{deGennes79}, in agreement with the well known
result that in 3d solutions the fluctuations relax via the self
diffusion of the polymer chain and not the mechanism outlined
above. In 2d relaxation by reptation is hindered by an effective
confinement provided by neighboring chains.

In summary, it has been shown that the thermal concentration
fluctuations in the semi-dilute regime of polymer monolayers
reported in the present paper and in ref.~\cite{monroy99} have a
slow mode of decay with a timescale that had not been previously
considered and that describes a specifically 2d process.

\begin{acknowledgments}
We thank  E.M.Terentjev for very useful comments and discussions.
\end{acknowledgments}

\bibliography{thesisv4}

\end{document}